# CLEAR: Coverage-based Limiting-cell Experiment Analysis for RNA-seq

Logan A Walker[1,2], Michael G Sovic[2], Chi-Ling Chiang[2,3], Eileen Hu[2,3], Jiyeon K Denninger[4], Xi Chen[2], Elizabeth D Kirby[4], John C Byrd[2,3], Natarajan Muthusamy[2,3], Ralf Bundschuh[1,3,5,6]*, & Pearlly Yan[2,3]*

[1]Department of Physics, College of Arts and Sciences
[2]The Ohio State University Comprehensive Cancer Center
[3]Division of Hematology, Department of Internal Medicine, College of Medicine
[4]Department of Psychology, College of Arts and Sciences
[5]Department of Chemistry & Biochemistry, College of Arts and Sciences
[6]Center for RNA Biology, The Ohio State University, Columbus, OH

* To whom correspondence should be addressed. Tel: +1 614 688 3978 and +1 614 685 9164; Email: bundschuh@mps.ohio-state.edu and Pearlly.Yan@osumc.edu

**ABSTRACT**

Direct cDNA preamplification protocols developed for single-cell RNA-seq have enabled transcriptome profiling of precious clinical samples and rare cells without sample pooling or RNA extraction. Currently, there is no algorithm optimized to reveal and remove noisy transcripts in limiting-cell RNA-seq (lcRNA-seq) data for downstream analyses. Herein, we present CLEAR, a workflow that identifies reliably quantifiable transcripts in lcRNA-seq data for differentially expressed gene (DEG) analysis. Libraries at three input amounts of FACS-derived CD5+ and CD5- cells from a chronic lymphocytic leukemia patient were used to develop CLEAR. When using CLEAR transcripts vs. using all transcripts, downstream analyses revealed more shared transcripts across different input RNA amounts, improved Principal Component Analysis (PCA) separation, and yielded more DEGs between cell types. As proof-of-principle, CLEAR was applied to an in-house lcRNA-seq dataset and two public datasets. When imputation is used, CLEAR is also adaptable to large clinical studies and for single cell analyses.

Deep sequencing of transcriptomes (RNA-seq) provides important insights into biology and disease. Bulk RNA-seq requires hundreds of thousands of cells. The resultant transcriptomic profile is therefore the average of cells at different transcriptomic states or even different cell types within the same tissues (e.g., infiltrating immune cells or normal cells in tumor samples). With the discovery of new genes and splice junctions in the first single-cell RNA-seq (scRNA-seq) study [1], researchers realized the need to profile single-cell transcriptomes. Coinciding with this intense interest is the development of diverse approaches to perform scRNA-seq, which have been summarized in recent reviews [2–5]. Improvements in reagents that enable full-length transcriptome profiling by direct global amplification at the single-cell level (e.g., SMART-seq [6–8], Quartz-Seq [9], and the 'Tang' method [1]) have also enabled direct amplification for the analysis of groups of 10's-100's cells, i.e., limiting-cell RNA-seq (lcRNA-seq).

The advantages of direct amplification are manifold. First, it lowers the barrier in identifying differentially expressed genes (DEGs) in rare cell populations. Until recently, researchers had to pool cells from multiple samples to extract quantifiable amounts of total RNA for library generation. With direct cDNA preamplification, the number of cells required for successful library preparation dropped below 100 cells, a level often achievable from just one sample. Also, quantifying RNA accurately below 250 pg/µL is challenging. Direct cDNA preamplification eliminates this need by using cell counts (e.g., from fluorescence activated cell sorting (FACS) or laser capture) instead of RNA mass to standardize input amounts between experimental groups. Second, direct cDNA preamplification greatly preserves transcript quality by quickly transforming labile RNA into stable cDNA, as degradation associated with extraction can be significant. Third, direct amplification of enriched cells deposited into well-plates allows the incorporation of nanoliter microfluidics to deliver/mix reagents and templates quickly. This further preserves RNA integrity.

Even with these advances, systems noise associated with transcript degradation is inevitable and requires computational solutions, especially if large numbers of replicates are not feasible. Current publications on lcRNA-seq data fall into two categories: cell-pool samples as part of method development or scRNA-seq methodology comparisons [6,10–12] and ultralow amounts of extracted RNA as input for RNA-seq library preparation [13–15]. Analysis workflows for bulk RNA-seq and scRNA-seq data are distinct as each approach addresses a different research question. The goal of bulk RNA-seq is to identify differences in transcriptomic profiles between treatment groups, whereas the goal of scRNA-seq is to characterize cell subpopulations in tissues or bulk cells. In this regard, the aim of lcRNA-seq experiments is like that of bulk RNA-seq experiments, whereas their data quality (e.g., prevalence of zero count genes or 'dropout rate' [10,16,17]) is similar to that of scRNA-seq. Therefore, statistical methods often used for between-group comparisons in bulk RNA-seq studies, such as the negative binomial distribution-based test in DESeq2 [18], should not be used for lcRNA-seq data without modifications because they are susceptible to zero-count artifacts. On the other hand, the myriad of tools [19–28] for analyzing scRNA-seq data are tuned to work with high variabilities (e.g., true biological variations in single cells or variabilities due to cDNA preamplification) but require large numbers of replicates not possible in lcRNA-seq studies.

Here, we describe CLEAR, a computational preprocessing approach for between-group comparisons of lcRNA-seq experiments. Designed for low numbers of replicates, CLEAR focuses on identifying robust transcripts with even read coverage for downstream analysis to control for data noise. It examines the noise pattern of individual samples to identify 'reliably quantifiable' transcripts for maximal signal and minimal noise. CLEAR transcripts common to replicates across two comparison groups will be used for subsequent analyses. Using a dataset derived from the same RNA stock but at dilutions spanning typical lcRNA-seq inputs, we show that CLEAR greatly improves similarity between results from three input RNA levels. In two public datasets, we demonstrate that the numbers and dispersion patterns of CLEAR transcripts yield a novel way to evaluate library qualities. In an in-house murine neural cell lcRNA-seq study, utilizing CLEAR transcripts significantly improves cell type separations by Principal Component Analysis (PCA) and validations of cell phenotype markers.

## RESULTS

**Developing CLEAR**

The experimental design for developing CLEAR is presented in Figure 1A. Our goal is to detect differences between two biological conditions using low input amounts and few replicates. Total RNA was extracted from FACS-derived primary CD5$^+$CD19$^+$ (CD5+) and CD5$^-$CD19$^+$ (CD5-) cells from a chronic lymphocytic leukemia (CLL) patient, thereby representing data quality typical of a clinical study. The RNA was diluted to levels found in typical lcRNA-seq projects with three replicates per input mass (10-, 100-, and 1,000-pg) and cell type. Although CLEAR is targeted for between-group comparisons in rare/limiting cell transcriptomes, sample inputs used for its development were from extracted RNA and not from cell pools. The overall design rationale is as follows: 1) To track the efficacy of transcripts selected by CLEAR in downstream DESeq2 [18] analysis, the composition of input RNA had to be the same for all dilutions so that the only difference within the same cell type was the input amount; 2) The 10- and 100-pg input RNA levels approximate the RNA in ~5 and ~50 immune cells, respectively. The 1,000-pg input level represents a 'gold standard' to compare the 10- and 100-pg to, as systems noise from this RNA amount is considered acceptable [10].

**LcRNA-seq data quality from serial dilutions of the same RNA stock strongly depends on input mass and yields dissimilar DEGs when processed with a standard RNA-seq analysis pipeline**

Average read depth was 15.5 million clusters, which is typical of bulk mRNA-seq experiments. Post-alignment quality assessment was used to illustrate the effect of input RNA amounts on data quality. As expected, quality parameters correlate inversely with RNA inputs (Figure S1), with the 10- and 100-pg samples yielding less transcriptomic information than the 1,000-pg samples despite having similar sequencing depths.

To examine the effect of technical noise associated with lcRNA-seq experiments, we performed DEG analysis using a standard bulk cell RNA-seq workflow on the 10-, 100- and 1,000-pg replicates without consideration of noise structure from signal dropouts and RNA degradation. As expected, the overall transcript variabilities, depicted in Figure S2 as scatterplots between replicates of the two cell types, were highest in the 10-pg input RNA replicates. Within all samples, variabilities were highest in the low- to medium-expressed transcripts while the expression profiles converged at high-expressed transcripts.

Next, we performed DEG analysis using DESeq2 workflow without modifications [18]. DEG counts for the CD5+ vs. CD5- comparison were: 1,000-pg: 2,996, 100-pg: 744, and 10-pg: 898 (Figure 1B). This outcome is unexpected as the statistical power to discern DEGs should increase with increasing RNA input mass. Of the DEGs identified between the two cell types, few were shared between any two input levels (Figure 1C). This is also unexpected because the three input levels were diluted from the same RNA extraction stocks. In Figure 1D, we present PCA plots for cell type comparisons at the three input levels. At the 10-pg input amount, the gene expression profiles from the three CD5- replicates were vastly different from each other resulting in no separation between the cell types along either Principal Component 1 (PC1) or PC2. At the 100- and 1,000-pg levels, the transcriptome profiles distinguished the two cell types, with PC1 accounting for 26.3% and 74.6% of the variance, respectively.

Taken together, the transcriptome profiles, the DEG dissimilarities and quantities, and the inconclusive PCA outcome reveal a high degree of systems noise at the 10-pg input level. Though systems noise was reduced at the 100- and 1,000-pg input amount, still relatively few shared DEGs were identified at these two levels. As such, lcRNA-seq data clearly require custom analysis to control for systems noise.

**Individual gene coverage profiles can guide selection of robust and reliably quantifiable transcripts in high noise conditions**

It is known that systems noise associated with lcRNA-seq data are highest in lowly-expressed transcripts and manifested as large amounts of duplicated reads [14,29]. As our aim is to provide a DEG analysis workflow for studies/samples with degraded RNA (e.g., multi-step cell-type enrichment from stored samples), low

RNA content (e.g., <100 immune cells), and few replicates, we elect to identify reliable transcripts in each replicate within each experimental group for between-group comparisons. The CLEAR workflow is presented in Figure 2A and in Methods.

In Figure 2B, we illustrate CLEAR transcript evaluation using coverage profiles from three genes in a 10-pg CD5+ sample. The first, GAPDH, is a highly-expressed housekeeping gene. This group of genes has uniform 5' to 3' read coverage, similar to coverage profiles in bulk RNA-seq. The second gene, RPS7, is in the last accepted CLEAR transcript bin. At this point, coverage profiles switch from an acceptable unimodality centering at the middle of the transcript to an unacceptable amount of bi-modality with region(s) of read pile-ups and region(s) with low/no coverage. Lastly, DDAH2 is below the CLEAR threshold. Here, the profile is dominated by regions with drastically different coverage. These profiles are prone to erroneous gene expression quantifications and false DEGs.

To illustrate the changes in coverage profiles, we plot the transcript coverage profile mean $\mu_i$ from CD5+ samples (one replicate per input level) in Figures 2C and 2D as scatter- and violin plots, respectively. The 100- and 1,000-pg samples have even read distributions ($\mu_i \approx 0$) for the 6,000 highest expressed transcripts while the 10-pg transcripts experience unevenness after the 2,000$^{th}$ highest expressed transcript. The red line signifies the transition of transcripts passing CLEAR to transcripts too noisy for DEG analysis. Due to higher input amounts in the 100- and 1,000-pg samples, their acceptable-to-unacceptable transitions occur at lower expression levels not shown in the plots (see discussion below).

**CLEAR selection increases shared DEGs between CD5+ and CD5- samples at all RNA input levels and improves cell type separation at the 10-pg level**

Earlier in this report (Figure 1B-D; Figure S2), we presented the impact of systems noise associated with lcRNA-seq libraries. In Figures 3 and S3-S4, we present evidence for the transformative power of removing noisy transcripts from lcRNA-seq DEG analysis. First, we examine the impact of CLEAR on the number of robust transcripts per input level. The red histogram (Figure 3A, *right* plot) shows 12,246 shared transcripts passing CLEAR in all 1,000-pg samples (three replicates/cell type). The number drops to 4,983 for the 100-pg samples and to 306 for the 10-pg samples. The high number of shared CLEAR transcripts among the 100- and 1,000-pg samples signifies low system noise, while few CLEAR transcripts are shared among the 10-pg samples. This highlights that input amounts ⩽5 cells are susceptible to systems noise. The same trends are observed when the shared CLEAR transcripts are investigated separately by cell type (Figure S3).

The effect of CLEAR on replicate-to-replicate comparisons is presented in Figure S4. The main takeaway is that CLEAR removes a large number of lowly-expressed transcripts from the 10-pg samples leaving highly-expressed transcripts for downstream analyses. The amount of retained transcripts increases with increasing RNA input. For the 1,000-pg samples, we observe transcripts from the entire transcription range passing CLEAR. This supports ours and others' postulation [10] that the molecular complexity in 1 ng RNA approaches that found in bulk RNA-seq. We note subtle differences between each comparison pair, illustrating that CLEAR assesses samples independently to derive non-static cutoff values.

The Venn diagrams in Figure 3B show the overlaps of CLEAR transcripts (depicted by the red bars in Figure 3A) common to all 6 samples (3 replicates/cell type) per input level. With CLEAR successfully excluding noisy transcripts from these lcRNA-seq data, nearly all of the shared CLEAR transcripts at each input level are contained in the shared CLEAR transcripts at the respective higher input level.

Similarly, CLEAR pre-selection positively impacts the DEG analysis between cell types (Figures 3C and 3D). In contrast to using all transcripts (Figure 1B; also Figure 3C *inset*), DEGs are now lowest (n=3) in the 10-pg comparison, intermediate (n=189) in the 100-pg comparison, and largest (n=2,826) in the 1,000-pg comparison. This trend is expected as power to detect DEGs increases with increasing RNA input. Also, CLEAR pre-selection alters the DEG overlap shown in Figure 1C. When only CLEAR transcripts are used (*bottom* Figure 3D), the 3 DEGs at the 10-pg inputs are: CD69 (found in all comparisons); SRSF11 (also found in the 1,000-pg comparison); and HINT1 (only found in the 10-pg comparison). Of the 189 DEGs at the 100-pg input level, a majority (n=135) is shared with the 1,000-pg input. These overlaps far exceed those using all transcripts (Figure 1C and *top* Figure 3D).

Finally, applying CLEAR also improves PCA segregation of cell types (Figure 3E). In contrast to results from data without CLEAR selection (Figure 1D), the CD5+ replicates are now separated from the CD5- replicates along PC1 at all input levels (Figure 3E) instead of just at the 100- and 1,000-pg levels as in Figure 1D.

To conclude, we remind readers that the dilution samples were prepared from the same RNA stock, the library generation was performed using the same reagents, by the same scientist, at the same time, and sequenced on the same flow cell. The dramatic differences in data quality observed in Figure 2C - 2D are mainly associated with the input RNA amount. While the traditional analysis shown in Figure 1 is severely hampered by the amount of systems noise in lcRNA-seq data, implementation of CLEAR results in overlapping and convergent DEGs at all input levels.

**The number of CLEAR transcripts correlates with authors' quality measures in public scRNA-seq data**

To evaluate CLEAR's utility in identifying robust transcripts in published ultralow-input full-length RNA-seq data, we selected an scRNA-seq dataset based on the Fluidigm/SMART-seq approach. Ilicic et al. [22] provided detailed quality characteristics on captured single cells. We selected the mouse embryonic stem cells (576 cells grown in 2i or alternative 2i media) to compare CLEAR's assessments of the captured cells vs. the authors' quality calls of 'Good' and 'Empty/No capture'. In Figure S5A, we present violin plots illustrating $\mu_i$ from scRNA-seq data from a 'Good' cell. We note that the coverage bifurcations observed in our lcRNA-seq data are also observed in the scRNA-seq data. In Figure S5B, we provide CLEAR transcript count distributions in the two quality-groups. For the 521 'Good' cells, they have similar CLEAR transcript counts with mean of 4,098. The 'Empty' cells, however, have widespread CLEAR transcript counts with mean at 651 and with 10 of 15 cells having 0. As depicted, the mean value is skewed by two outliers, potentially due to ambient RNA from cell bursting prior to imaging.

**CLEAR transcript numbers are similar between studies with similar input RNA mass**

Next, we applied CLEAR to SMART-Seq data from Bhargava et al. [14] prepared from serial dilution (25-, 50-, 100- and 1,000-pg) of polyA-selected mRNA from control and Activin A-treated mouse embryonic stem cells. Again, we note that the read coverage bifurcations observed in our lcRNA-seq data also occurred here (Figure S5C). Figure S5D presents CLEAR transcript counts for the four samples (control and treated groups, two replicates each) at all input levels. As expected, CLEAR transcript counts increase with increasing mRNA inputs. To compare CLEAR transcript counts in the Bhargava et al. [14] data to our CD5+/CD5- data, one should note the difference in type of input material (polyA-selected mRNA in the former and total RNA in the latter) and associated differences in sample preparation chemistry. We apply a conservative adjustment factor of 10 [30] between total RNA and mRNA. As such, our 1,000-pg total RNA 'gold standard level' is comparable to Bhargava et al.'s [14] 100-pg mRNA input. Indeed, the Bhargava et al. [14] 100-pg mRNA libraries produced from Activin A stimulated cells had similar shared CLEAR transcript counts (mean, 14,940) to our 1,000-pg CD5+/CD5- replicates (mean, 14,717). It is worth noting, however, that the unstimulated control samples from the Bhargava study produced many fewer CLEAR transcripts (mean, 6,735), demonstrating the ability of CLEAR to distinguish active- and quiescent state between samples. At the intermediate input level of 100-pg total RNA, the closest corresponding level for the Bhargava et al. [14] study would be 10-pg mRNA. Since their lowest input level was 25-pg, it makes sense that their average CLEAR transcript counts of 6,408 are similar to our average transcript counts of 8,420 at 100-pg input. Due to the higher RNA input amount, the data in the Bhargava et al. [14] study was not challenging enough to highlight CLEAR's ability to improve DEG and PCA analyses between the two biological groups, in agreement with the authors' original assessment.

**CLEAR improves group separation of murine neural cell types and recapitulates genes known to be differentially expressed between them**

To demonstrate the applicability of CLEAR in a biological system, we applied it to lcRNA-seq data derived from three cell types in murine hippocampal dentate gyrus (DG), a highly-studied neurogenic niche associated with memory and cognitive function [31]. This region is of particular interest because resident stem and glial cells show robust morphologic and proteomic responses to a variety of injuries such as trauma and seizures [32,33]. However, the stem cells and local glial populations are small cell populations embedded in a heterogeneous niche thereby perfect candidates for FACS-derived lcRNA-seq analysis.

The study schematics are shown in Figure 4A and the variability of CLEAR transcript counts in Figure 4B. The data reveal that neural progenitor cells had the highest number of CLEAR transcripts, followed by neural stem cells and then astrocytes. While this trend may reflect differences in cell size and RNA content, it may also reveal the tolerance of each cell type toward the cell dissociation and enrichment conditions. For example, astrocytes possess a complex branching of fine processes extending from their cell body that could be damaged during the mechanical dissociation and flow sorting. Furthermore, they are intrinsically reactive to changes in the external environment [34] such as the chemical dissociation step and the FACS media required by the Clontech SMARTer kit (phosphate-buffered saline devoid of fetal bovine serum, $Ca^{2+}$, and $Mg^{2+}$).

Despite suboptimal cDNA quality from the astrocytes, the application of CLEAR greatly enhanced cell-type separations when compared to using all available transcripts (Figure 4C). When all transcripts were used, PCA plots show clear separation between stem cells and progenitors (third panel on *top*), but the lower transcript quality in astrocytes renders the discrimination between astrocytes and stem cells unclear (first panel on *top*). The astrocyte transcript quality does not impact its separation from the progenitors due to large differences in transcript profiles between these two cell types (second panel on *top*). Upon application of CLEAR, astrocytes and stem cells become well separated (first panel on *bottom*) while retaining separation between astrocytes and progenitors (second panel on *bottom*).

Genes known to be differentially expressed in the three cell types (Table 1) are used to confirm the efficacies of the enrichment strategies and of the CLEAR analysis. Of the transcripts listed in Table 1, only the four shown in the top section and in Figure 4D passed CLEAR in all three cell types. Genes that failed CLEAR are presented in Figure S6. Glial fibrillary acidic protein (Gfap) and SRY-Box 9 (Sox9) are known to be higher in astrocytes and stem cells than in progenitors, which we also found ($q < 0.01$ and $0.0001$, respectively). Glutamate ionotropic receptor NMDA type subunit 2C (Grin2c) and HOP homeobox (Hopx) are known to be different between stem cells and astrocytes, with Hopx higher in stem cells and Grin2c higher in astrocytes. Our findings confirmed both trends ($q < 0.01$ and $q < 0.001$, respectively). In addition, our data show that the expressions of Hopx and Grin2c are significantly higher in stem cells and astrocytes than in progenitors ($q < 0.0001$ for both). Finally, the expression of fatty acid binding protein 7 (Fabp7) is known to be higher in progenitors and stem cells relative to astrocytes as confirmed in our experiment ($q < 0.01$ and $q < 0.05$, respectively).

For genes that do not pass CLEAR, the picture is more nuanced. We examine normalized count plots for a few of the known discriminating transcripts (Figure S6). Inhibitor of DNA binding 4 (Id4) and SRY-Box 2 (Sox2) are known to have similar transcript levels in astrocytes and stem cells but higher than those in progenitors, which is what we found. Minichromosome maintenance complex component 2 (Mcm2), doublecortin (Dcx), eomesodermin (Eomes), proliferating cell nuclear antigen (Pcna), neuronal differentiation 1 (Neurod1), and neurogenin 2 (Neurog2) have higher expressions in the progenitors [35–40], which we also find. However, nestin (Nes) is expected to be expressed in both stem cells and progenitors, but is only observed in the latter. Without CLEAR, Nes would be called as significantly higher in progenitors ($q \leq 0.01$). With CLEAR, Nes fails its criteria thereby avoiding a false finding. Taken together, our dataset recapitulates and reinforces relative gene expression patterns reported for neural stem cells, progenitors, and astrocytes and shows that staining, sorting, and lcRNA-seq with CLEAR can be used to distinguish these cell types.

**DISCUSSION**

Total RNA input amount for lcRNA-seq (e.g., 10 - 1,000-pg from 1 - 100 cells) is closer to amounts found in scRNA-seq studies (1 - 10-pg) than to bulk RNA-seq studies (100 - 200-ng). Due to the minute RNA input amount in single-cell/limiting-cell transcriptomic studies, the library generation process, with the exception of single-molecule sequencing, is preceded by global cDNA preamplification. Researchers have observed artefacts such as a large proportion of the sequencing reads being dominated by a small number of transcripts [14], excessive transcripts with zero read counts [41], high variabilities between and within replicates of sample groups [42], distortion towards shorter transcripts [43], and higher variances at lower biological abundances [10,11]. Together, these artefacts result in noisy sc/lcRNA-seq data that challenge assumptions fundamental to bulk RNA-seq analysis approaches and render them unsuitable for direct application to these data [16]. Existing strategies to overcome challenges associated with noisy data to uncover biological differences in single-cell studies include: 1) identifying low-quality, high-noise samples and removing them from downstream analyses [22]; 2) applying transcript normalization [16,24,26,44]; 3) incorporating ERCC spike-in control [42,45]; 4) utilizing median absolute deviation (more resistant to outliers) to characterize statistical dispersion [12]; 5) integrating UMIs to track transcripts by molecular counts [46]. Yet, due to the often low number of replicates and the goal of identifying DEGs between sample groups, the strategies for scRNA-seq data just described are not appropriate or adequate for lcRNA-seq data.

In this paper, we describe CLEAR, an objective approach to identify transcripts in lcRNA-seq with acceptable systems noise for downstream analyses. CLEAR requires transcripts to be reliably quantifiable in all replicates within and across comparison groups. This criterion is made possible due to the RNA quality preservation effects of direct cDNA preamplification reagents and library generation using nanoliter-microfluidics devices as they increase even read coverage in lcRNA-seq transcripts.

To our knowledge, preprocessing lcRNA-seq data using CLEAR followed by DEG analysis is unique and unlike approaches used by others in similar studies. For example, Shanker et al. [13] performed Pearson correlation comparisons between serial dilutions of input RNA vs. their 1 µg input control using $log_2$ of median RPKM for all genes having ≥ 2 reads coverage. Bhargava et al. [14] performed DESeq without prior data preprocessing to compare the performance between three library preparation methods as well as within serial dilutions of mRNA derived from two mouse embryonic stem cell culture conditions. Liu et al. [47] utilized relative expression orderings to harmonize clinical transcription signatures between low-input and bulk RNA-seq libraries. In contrast, CLEAR comprehensively screens all transcripts in all replicates within an lcRNA-seq study for genes with even coverage for downstream analyses. The combination of characterizing each transcript by its read distribution mean $\mu_i$ followed by cutoff selection defined by expression across the full transcript distribution is deliberate. For any given transcript, even one with low systems noise, there are many reasons why its mean $\mu_i$ could deviate from zero: 1) transposase-based library generation methods [48] less effectively 'tagment' the 5'-end of transcripts (3'-ends are less affected likely due to the presence of poly(A) tails), affecting 5'-end read coverage; 2) transcript isoforms skew the read distribution along the length of the transcript, especially when prominent isoforms are not correctly annotated [17]; 3) presence of truncated, polyadenylated RNA fragments/intermediates part way through the RNA decay pathway [49]; 4) presence of RNA fragments with alternative polyadenylation sites [17,50,51]. Thus, a non-zero $\mu_i$ alone does not signify a noisy transcript. Conversely, read coverage profiles of some of the noisy transcripts can become random and $\mu_i$ can approach zero by chance. Simply keeping transcripts with a mean $\mu_i$ close to zero would result in some valid transcripts being excluded and some noisy transcripts being included. By characterizing the entire distribution of $\mu_i$ over a defined range of expression levels, outliers due to the above-mentioned effects do not impact the classification of that particular expression bin. In this manner, CLEAR adapts to the quality of each sample to assess systems noise, an approach more precise than using an arbitrary cutoff.

In this report, we have selected two sets of public data [14,22] and one set of in-house data to illustrate the utility of CLEAR. One of the goals for using external datasets was to ascertain that the mean read/fragment distribution phenomenon observed in the CLEAR development data is also present in sc- and lcRNA-seq libraries derived from Fluidigm C1 IFCs and manual SMART-seq, respectively. As suspected, the violin plots from the public data (Figures S6A, and S6C) illustrate the distribution of $\mu_i$ transitions from an acceptable distribution centered around 0 to an unacceptably broad/bimodal distribution.

Another goal was to evaluate the ability of CLEAR to identify noisy transcripts in a broad range of datasets. The application of CLEAR to Ilicic et al. scRNA-seq data [22] revealed association between cell integrity and the resultant data quality. Figure S5B displays violin plots of the dispersion of CLEAR transcripts which recapitulated the authors' microscopy assessments of intact 'single cells' and 'no cell'. When applied to Bhargava et al. data [14,22], CLEAR identified increasing amounts of 'passed' transcripts with increasing mRNA input mass (Figure S5D) which we also observe in the CD5+/CD5- data. Together, these assessments confirm that CLEAR can independently reveal sample/transcript quality and identify noisy data associated with empty wells (Ilicic et al. data [14,22]) and increases in data quality with increasing levels of RNA input (Bhargava et al. data [14,22]). As yet another example of CLEAR's performance, we incorporated a proof-of-principle experiment with cells isolated from the neurogenic niche in murine hippocampus. Traditionally, this study has been hampered by limited numbers of stem cells and progenitors present in murine DG. By coupling preamplification-based lcRNA-seq with CLEAR, DEGs between these cell types can be achieved with only one mouse brain per biological replicate.

When comparing different biological groups it is possible that a gene is highly expressed in one but not in the other. An example of this is depicted in the Eomes gene in Figure S6. It passes CLEAR in the progenitors but fails in the other cell types. While true DEG analysis is not possible for such genes, one can still use the CLEAR cutoff expression level (the count of the lowest transcript passing CLEAR for a particular sample) to report a bound on the fold change of such a gene consistent with the CLEAR analysis. For example, DESeq2 reports a $\log_2$ fold change of 8.97 when comparing the expression of Eomes of progenitors to astrocytes. However, the 'bound' procedure suggests a more modest difference of 1.48, when substituting the threshold for the featureCounts-reported expression value. The latter analysis is likely more reliable than the higher fold change reported by DESeq2 alone due to the presence of signal dropouts when all transcripts are used.

Although CLEAR can be applied to scRNA-seq data just as effectively as lcRNA-seq data (Figure S5A-B, results from analyzing Ilicic et al. [22] scRNA-seq data using CLEAR), the low complexity and highly variable nature of scRNA-seq data would typically result in insufficient CLEAR transcripts for meaningful DEG analysis, especially when requiring transcripts to pass CLEAR criterion in all (usually of a very large number) cells. The incorporation of imputation would relax the CLEAR criterion thereby rendering it useful in scRNA-seq data.

In summary, utilizing a direct cDNA amplification approach, the lcRNA-seq strategy allows researchers to perform global transcriptome profiling and identify DEGs using a few cells. We develop the CLEAR algorithm around two prevalent phenomena in ultralow input RNA-seq data: 1) systems noise due to global preamplification of low RNA amount is more prevalent in medium- to lowly expressed transcripts; 2) the read coverage profile of noisy transcripts contain discernible coverage gaps instead of even coverage along the transcript length. By gauging the average read coverage profile mean $\mu_i$ in transcripts binned by gene expression levels, CLEAR is able to systematically identify transcripts with sufficient integrity in all replicates across comparison groups for downstream analyses. Lastly, we note that by displaying $\mu_i$ as violin plots, CLEAR is useful as a visual QC tool to eliminate or set aside data from scRNA-seq samples that fail at the cell enrichment stage or during library generation. In all, we highlighted the functionalities CLEAR brings to sc/lcRNA-seq data preprocessing and data quality visualization and hope to spark additional computational and statistical development in this area.

**METHODS**

*CLL Patient Sample Acquisition*
A chronic lymphocytic leukemia (CLL) patient sample was obtained from the Leukemia Tissue Bank (LTB), a shared resource of the NCI-funded OSU Comprehensive Cancer Center. The sample was obtained following written informed consent in accordance with the Declaration of Helsinki and under a protocol reviewed and approved by the Institutional Review Board of the Ohio State University. The patient had CLL as defined by the IWCLL 2008 criteria. The patient's white blood cells were isolated by Ficoll density gradient centrifugation (Ficoll-Paque Plus, Amersham Biosciences, Little Chalfont, UK) and samples were banked at -180°C in liquid nitrogen. Frozen cells were thawed and washed with RPMI 1640 media (Gibco,

Life Technologies, Grand Island, NY, USA) and resuspended at $5 \times 10^6$ cells/mL in complete medium containing 10% fetal bovine serum (FBS) (Sigma, St Louis, MO, USA), 2 mM l-glutamine, penicillin (100U/mL), and streptomycin (100 μg/ mL) (Gibco).

*Animals for Neural Cell Type Analysis*
All procedures involving animals were approved by the Ohio State Institutional Animal Care and Use Committee in accordance with institutional and national guidelines. Nestin-GFP mice were provided by Grigori Enikolopov at Cold Spring Harbor Laboratory[52]. All mice were housed in a 12 hour light-dark cycle with food and water ad libitum. For isolation of the dentate gyrus (DG), adult mice (6-9 wk old) were anesthetized with an intraperitoneal injection of ketamine (87.5 mg/kg) and xylazine (12.5 mg/kg) before perfusion with PBS. Following perfusion, the brain was removed and placed in cold Neurobasal A medium (Gibco 10-888-022) on ice. After bisecting the brain along the midsagittal line, the cerebellum and diencephalic structures were removed to expose the hippocampus. Under a dissection microscope (Zeiss), the DG was excised using a beveled syringe needle and placed in ice cold PBS without calcium or magnesium (Gibco 10-010-049). DGs from mice were first mechanically dissociated with sterile scalpel blades and then enzymatically dissociated with a pre-warmed papain (Roche 10108014001)/dispase (Stem Cell Technologies 07913)/DNase (Stem Cell Technologies NC9007308) (PDD) cocktail at 37°C for 20 min. Afterwards, the tissue was again mechanically disrupted by trituration for 1 min. Dissociated cells were collected by centrifugation at 500g for 5 min.

*Fluorescence activated cell sorting (FACS)*
For the human sample all cells were stained and sorted by FACS Aria (BD Biosciences, San Jose, CA, USA). Live CLL B cells (CD5+CD19+) and normal B cells (CD5-CD19+) were sorted from the patient sample. Briefly, cells for FACS were resuspended in PBS without calcium/magnesium and filtered through a 35μm nylon filter and then stained for PE-conjugated CD45 (HI30), PerCp Cy-5.5-conjugated CD19 (HIB19), and Alexa Fluor 700-conjugated CD3 (UCHT1) monoclonal antibodies. Nonviable cells were excluded by the LIVE/DEAD Fixable Near-IR Dead Cell Stain Kit (Life Technologies, Carlsbad, CA, USA). Appropriate fluorescence minus one controls were used to determine nonspecific background staining. Single-cell gates were used to exclude the possibility of doublet cells. The FACS parameter diagrams for this process are available in Figure S7.

For the Nestin-GFP mouse samples, cells were resuspended in PBS without calcium/magnesium and filtered through a 35μm nylon filter before staining with the following antibodies: O4-APC (1:100, R&D FAB1326A), O1-eFluor660 (1:100, Thermo Fisher/eBioscience, Pittsburgh, PA, 5065082), GLAST-PE (1:50, Miltenyi Biotec, Bergisch Gladbach, Germany, 130-098-804), CD45-APC (1:100, BD Biosciences, Franklin Lakes, NJ, 561018), CD31-APC (1:100, BD Biosciences, Franklin Lakes, NJ, 561814). Cells were incubated with antibodies on ice for 30 minutes. During the last 10 minutes of staining, Hoechst dye (1:10,000, Thermo Fisher, Pittsburgh, PA, 33342) was added for live/dead discrimination. All cells were washed twice following staining and immediately sorted as stem, progenitor, or astrocyte populations based on fluorescent markers with the FACSAria III (BD Biosciences, Franklin Lakes, NJ). CD31, CD45, O1, and O4 negative live cells were designated as stem cells if double positive for GLAST and GFP, progenitors if only GFP positive, and astrocytes if only GLAST positive. For cells in the limited cell number study, 50 cells were sorted into 96 well format plate for downstream lcRNA-seq library generation.

*Total RNA Extraction*
Total RNA extraction was performed using Trizol reagent (Invitrogen, Carlsbad, CA). Briefly, approximately two million FACS-derived CD5+ and CD5- cells were separately sorted into 1.7 ml microcentrifuge tubes. Excess buffer was removed by centrifugation. Trizol reagent was added to cell pellets and the extraction protocol recommended by Invitrogen was followed. Total RNA was precipitated with 10μg glycogen (Qiagen, Hilden, Germany). The quality of the total RNA was assessed with the Agilent 2100 Bioanalyzer (Agilent, Inc., Santa Clara, CA) using total RNA Pico chip.

*Library Generation and Sample Sequencing*
Total CD5+ and CD5- RNA quantified using the Invitrogen Qubit RNA HS Assay kit (Invitrogen, Carlsbad, CA) was serially diluted to masses characteristic of single- and limiting-cell RNA-Seq (10-, 100-, and 1000-pg). The Clontech SMARTer v4 kit (Takara Bio USA, Inc., Mountain View, CA) was used for global preamplification of these serially diluted samples in triplicate and also for direct global preamplification in FACS-derived murine DG cell types prior to library generation in quadruplicates with the Nextera XT DNA Library Prep kit (Illumina, Inc., San Diego, CA). Samples were sequenced to a depth of 15 - 20 million 2x150 bp clusters on the Illumina HiSeq 4000 platform (Illumina, Inc., San Diego, CA).

*Publicly Available Data Retrieval*
Published data was retrieved in FASTQ format from the following accession numbers: ArrayExpress E-MTAB-2600 [53] (mouse embryonic stem cells from 2i and alternative 2i media) and GSE50856 [14] (SMART-seq samples).

*Data Preprocessing, Alignment, and Quantification*
Individual FASTQ files were trimmed for adapter sequences and filtered for a minimum quality score of Q20 using AdapterRemoval v2.2.0 [54]. Preliminary alignment using HISAT2 v2.0.6 [55] was performed to a composite reference of rRNA, mtDNA, and PhiX bacteriophage sequences obtained from NCBI RefSeq [56]. Reads aligning to these references were excluded in downstream analyses. Primary alignment was performed against the human genome reference GRCh38p7 or mouse genome reference GRCm38p4 using HISAT2. Gene expression values for genes described by the GENCODE [57,58] Gene Transfer Format (GTF) release 25 (human) or release M14 (mouse) were quantified using the featureCounts tool of the Subread package v1.5.1 [59] in unstranded mode.

*Sequencing and Alignment Quality Control*
Quality control was performed using a modification of our custom workflow QuaCRS [60]. In brief, aligned read quality was verified using RNA-SeQC [61] and RSeQC [62]. Parameters evaluated included the exonic rate (the percentage of reads aligning to exons), the intronic rate (the percentage of reads aligning to introns), and the duplication rate (the percentage of reads that were identified as PCR duplicates).

*Coverage Profiling*
Coverage depths across the aligned reference were calculated with a per-base resolution using the 'genomecov' utility of Bedtools v2.27.0 [63] in BedGraph format. It is imperative to utilize 'split output' mode to reduce the size of the output BedGraph files. These files are used as input into CLEAR.

*Calculation of Transcript $\mu_i$*
For each transcript *i* annotated in the NCBI RefSeq GRCh38 reference sourced from the UCSC Genome Browser [64], CLEAR calculates the transcript's $\mu_i$ parameter. This quantifies the distribution of the positional mean of the read distribution along that transcript between the 5' ($\mu_i = -1$) and the 3' ($\mu_i = +1$) ends (Equation 1):

$$\mu_i = \frac{2}{L_{\mathrm{transcript}}}\left( \frac{\sum_{k=0}^{L_{\mathrm{transcript}} - 1} \left( k\cdot d_k \right) }{\sum_{k=0}^{L_{\mathrm{transcript}} - 1} \left(d_k\right) } \right) - 1 \quad (1)$$

$\mu_i = \frac{2}{L_{\mathrm{transcript}}}\left( \frac{\sum_{k=0}^{L_{\mathrm{transcript}} - 1} \left( k\cdot d_k \right) }{\sum_{k=0}^{L_{\mathrm{transcript}} - 1} \left(d_k\right) } \right) - 1$

where $L_{transcript}$ is the length of a given transcript, and $d_k$ is the coverage of exonic locus *k* zero indexed and starting at the transcription start site.

*Determination of Analysis-Ready CLEAR Transcripts*
All transcripts quantified by featureCounts are sorted by overall length-normalized expression. Histograms, of $\mu_i$ values from 250 transcripts each, are collected and fit using the optimize module of the Python scipy package, to a double-beta distribution as described by Equation 2:

$$H\left( \frac{(1+x)(1-\frac{1}{2}(1+x))^{1+b}}{2B(2,2+b)} + \frac{(1+x)^{1+a}(1-\frac{1}{2}(1+x))}{2^{1+a}B(2+a,2)} \right) \quad (2)$$

$$H \left( \frac{(1+x)(1-\frac{1}{2}(1+x))^{1+b}}{2B(2, 2+b)} + \frac{(1+x)^{1+a}(1-\frac{1}{2}(1+x))}{2^{1+a} B(2+a,2)} \right)$$

where *H* is a normalization parameter fixed by the bin sizes, *B(q,p)* is the beta integral of *q* and *p*, and *x* is the bin location. The fitting parameters are *a* and *b*, which are each bounded to be non-negative. A value of 0 for both parameters represents a symmetric distribution and positive values represent progressively bimodal distributions to the positive (*a*), or negative (*b*) direction. These windows are advanced in units of 10 transcripts, reducing computational intensity. Once a window is found where *a* or *b* have a value greater than 2, the software exports transcripts with a measured expression higher than that bin's value for analysis. For multi-sample comparisons, transcripts are overlapped and only transcripts found in all samples are included for downstream analysis.

*CLEAR Visualizations*
A core component to the quality control of the CLEAR selection process is the visualization of $\mu_i$ to confirm that the characteristic bifurcation is observed. Violin plots are produced using code included with CLEAR which utilizes the matplotlib package in Python. Examples of this can be seen in Figures 2D, S5A, and S5C.

*DEG Comparisons*
Differentially expressed genes were called using DESeq2 [18] run on counts tables generated with featureCounts as described above. In all summarization figures, a false discovery rate (FDR) q-value of <0.05 was used as an inclusion criterion for DEGs.

*Principal Component Analysis*
Principal component analysis (PCA) was utilized to visualize differences between samples. All PCA plots were generated from counts tables that were size-normalized and r-log transformed after CLEAR selection using methods included with DESeq2. Each comparison was processed with the SciKitLearn [65] PCA implementation and plotted using custom scripts in Python.

*Data Availability*
All original sequencing files have been deposited to Gene Expression Omnibus (GEO) under accession numbers GSE115032 (human CD5+ and CD5- data) and GSE115033 (mouse neural data).

*Code Availability*
The reference Python code implementing CLEAR is available for download at https://github.com/rbundschuh/CLEAR under the GPL3 license.

**ACKNOWLEDGEMENTS**

This work was supported in part by The Ohio State University Comprehensive Cancer Center and the National Institutes of Health (NIH) [P30 CA016058 (Genomics Shared Resource)]; the Pelotonia Foundation [fellowships to L.A.W., C.C, E.H.]; the NIH [R50 CA211524-03 and U54 CA217297 to P.Y.]; a OSU Division of Hematology Seed Grant to P.Y. and N.M.; the Chronic Brain Injury and Discovery Themes at The Ohio State University [seed funding to E.K.]; and allocations of computation resources from the Ohio Supercomputer Center [66]. Funding for open access charge: OSU Division of Hematology.

**AUTHOR CONTRIBUTIONS**

Conceived and designed the experiments: L.A.W., E.D.K, J.C.B., N.M., R.B., P.Y. Performed the experiments: M.G.S., C-L.C., E.H., J.K.D., X.C. Analyzed the data: L.A.W. Wrote the paper: L.A.W., E.D.K, J.K.D., R.B., P.Y.

**ADDITIONAL INFORMATION**

Supplementary Information accompanies this paper at http://www.nature.com/naturecommunications.

**COMPETING FINANCIAL INTERESTS**

The authors declare that they do not have any competing interests.

**REFERENCES**

1. Tang, F. *et al.* mRNA-Seq whole-transcriptome analysis of a single cell. *Nat. Methods* **6**, 377–382 (2009).

2. Picelli, S. Single-cell RNA-sequencing: The future of genome biology is now. *RNA Biol.* **14**, 637–650 (2017).

3. Ziegenhain, C. *et al.* Comparative Analysis of Single-Cell RNA Sequencing Methods. *Mol. Cell* **65**, 631–643.e4 (2017).

4. Haque, A., Engel, J., Teichmann, S. A. & Lönnberg, T. A practical guide to single-cell RNA-sequencing for biomedical research and clinical applications. *Genome Med.* **9**, 75 (2017).

5. Gupta, I. *et al.* Single-cell isoform RNA sequencing characterizes isoforms in thousands of cerebellar cells. *Nat. Biotechnol.* (2018). doi:10.1038/nbt.4259

6. Ramsköld, D. *et al.* Full-length mRNA-Seq from single-cell levels of RNA and individual circulating tumor cells. *Nat. Biotechnol.* **30**, 777–782 (2012).

7. Picelli, S. *et al.* Smart-seq2 for sensitive full-length transcriptome profiling in single cells. *Nat. Methods* **10**, 1096–1098 (2013).

8. Picelli, S. *et al.* Full-length RNA-seq from single cells using Smart-seq2. *Nat. Protoc.* **9**, 171 (2014).

9. Sasagawa, Y. *et al.* Quartz-Seq: a highly reproducible and sensitive single-cell RNA sequencing method, reveals non-genetic gene-expression heterogeneity. *Genome Biol.* **14**, R31 (2013).

10. Marinov, G. K. *et al.* From single-cell to cell-pool transcriptomes: stochasticity in gene expression and RNA splicing. *Genome Res.* **24**, 496–510 (2014).

11. Streets, A. M. *et al.* Microfluidic single-cell whole-transcriptome sequencing. *Proc. Natl. Acad. Sci. U. S. A.* **111**, 7048–7053 (2014).

12. Wu, A. R. *et al.* Quantitative assessment of single-cell RNA-sequencing methods. *Nat. Methods* **11**, 41–46 (2014).

13. Shanker, S. *et al.* Evaluation of commercially available RNA amplification kits for RNA sequencing using very low input amounts of total RNA. *J. Biomol. Tech.* **26**, 4–18 (2015).

14. Bhargava, V., Head, S. R., Ordoukhanian, P., Mercola, M. & Subramaniam, S. Technical variations in

low-input RNA-seq methodologies. *Sci. Rep.* **4**, 3678 (2014).

15. Adiconis, X. *et al.* Comparative analysis of RNA sequencing methods for degraded or low-input samples. *Nat. Methods* **10**, 623–629 (2013).

16. Vallejos, C. A., Risso, D., Scialdone, A., Dudoit, S. & Marioni, J. C. Normalizing single-cell RNA sequencing data: challenges and opportunities. *Nat. Methods* **14**, 565–571 (2017).

17. Grün, D. & van Oudenaarden, A. Design and Analysis of Single-Cell Sequencing Experiments. *Cell* **163**, 799–810 (2015).

18. Love, M. I., Huber, W. & Anders, S. Moderated estimation of fold change and dispersion for RNA-seq data with DESeq2. *Genome Biol.* **15**, 550 (2014).

19. Trapnell, C. *et al.* The dynamics and regulators of cell fate decisions are revealed by pseudotemporal ordering of single cells. *Nat. Biotechnol.* **32**, 381–386 (2014).

20. McCarthy, D. J., Campbell, K. R., Lun, A. T. L. & Wills, Q. F. Scater: pre-processing, quality control, normalization and visualization of single-cell RNA-seq data in R. *Bioinformatics* **33**, 1179–1186 (2017).

21. Angerer, P. *et al.* destiny: diffusion maps for large-scale single-cell data in R. *Bioinformatics* **32**, 1241–1243 (2016).

22. Ilicic, T. *et al.* Classification of low quality cells from single-cell RNA-seq data. *Genome Biol.* **17**, 29 (2016).

23. Lin, P., Troup, M. & Ho, J. W. K. CIDR: Ultrafast and accurate clustering through imputation for single-cell RNA-seq data. *Genome Biol.* **18**, 59 (2017).

24. Finak, G. *et al.* MAST: a flexible statistical framework for assessing transcriptional changes and characterizing heterogeneity in single-cell RNA sequencing data. *Genome Biol.* **16**, 278 (2015).

25. Guo, M., Wang, H., Steven Potter, S., Whitsett, J. A. & Xu, Y. SINCERA: A Pipeline for Single-Cell RNA-Seq Profiling Analysis. *PLoS Comput. Biol.* **11**, e1004575 (2015).

26. Vallejos, C. A., Marioni, J. C. & Richardson, S. BASiCS: Bayesian Analysis of Single-Cell Sequencing Data. *PLoS Comput. Biol.* **11**, e1004333 (2015).

27. van Dijk, D. *et al.* Recovering Gene Interactions from Single-Cell Data Using Data Diffusion. *Cell* **174**, 716–729.e27 (2018).

28. Butler, A., Hoffman, P., Smibert, P., Papalexi, E. & Satija, R. Integrating single-cell transcriptomic data across different conditions, technologies, and species. *Nat. Biotechnol.* **36**, 411–420 (2018).

29. Islam, S. *et al.* Quantitative single-cell RNA-seq with unique molecular identifiers. *Nat. Methods* **11**, 163–166 (2014).

30. Warner, J. R. The economics of ribosome biosynthesis in yeast. *Trends Biochem. Sci.* **24**, 437–440 (1999).

31. Deng, W., Aimone, J. B. & Gage, F. H. New neurons and new memories: how does adult hippocampal neurogenesis affect learning and memory? *Nat. Rev. Neurosci.* **11**, 339 (2010).

32. Weston, N. M. & Sun, D. The Potential of Stem Cells in Treatment of Traumatic Brain Injury. *Curr. Neurol. Neurosci. Rep.* **18**, 1 (2018).

33. Kuruba, R., Hattiangady, B. & Shetty, A. K. Hippocampal neurogenesis and neural stem cells in temporal lobe epilepsy. *Epilepsy Behav.* **14**, 65–73 (2009).

34. Liddelow, S. A. & Barres, B. A. Reactive Astrocytes: Production, Function, and Therapeutic Potential. *Immunity* **46**, 957–967 (2017).

35. Artegiani, B. *et al.* A Single-Cell RNA Sequencing Study Reveals Cellular and Molecular Dynamics of the Hippocampal Neurogenic Niche. *Cell Rep.* **21**, 3271–3284 (2017).

36. Hochgerner, H., Zeisel, A., Lönnerberg, P. & Linnarsson, S. Conserved properties of dentate gyrus neurogenesis across postnatal development revealed by single-cell RNA sequencing. *Nat. Neurosci.* **21**, 290–299 (2018).

37. Dulken, B. W., Leeman, D. S., Boutet, S. C., Hebestreit, K. & Brunet, A. Single-Cell Transcriptomic Analysis Defines Heterogeneity and Transcriptional Dynamics in the Adult Neural Stem Cell Lineage. *Cell Rep.* **18**, 777–790 (2017).

38. Shin, J. *et al.* Single-Cell RNA-Seq with Waterfall Reveals Molecular Cascades underlying Adult Neurogenesis. *Cell Stem Cell* **17**, 360–372 (2015).

39. Shi, Z. *et al.* Single-cell transcriptomics reveals gene signatures and alterations associated with aging in distinct neural stem/progenitor cell subpopulations. *Protein Cell* **9**, 351–364 (2018).

40. Zhang, J. & Jiao, J. Molecular Biomarkers for Embryonic and Adult Neural Stem Cell and Neurogenesis. *Biomed Res. Int.* **2015**, (2015).

41. Lun, A. T. L., Bach, K. & Marioni, J. C. Pooling across cells to normalize single-cell RNA sequencing data with many zero counts. *Genome Biol.* **17**, 75 (2016).

42. Brennecke, P. *et al.* Accounting for technical noise in single-cell RNA-seq experiments. *Nat. Methods* **10**, 1093–1095 (2013).

43. Hayashi, T. *et al.* Single-cell full-length total RNA sequencing uncovers dynamics of recursive splicing and enhancer RNAs. *Nat. Commun.* **9**, 619 (2018).

44. Kharchenko, P. V., Silberstein, L. & Scadden, D. T. Bayesian approach to single-cell differential expression analysis. *Nat. Methods* **11**, 740–742 (2014).

45. Ding, N., Melloni, L., Zhang, H., Tian, X. & Poeppel, D. Cortical tracking of hierarchical linguistic structures in connected speech. *Nat. Neurosci.* **19**, 158 (2015).

46. Tung, P.-Y. *et al.* Batch effects and the effective design of single-cell gene expression studies. *Sci. Rep.* **7**, 39921 (2017).

47. Liu, H. *et al.* Robust transcriptional signatures for low-input RNA samples based on relative expression orderings. *BMC Genomics* **18**, 913 (2017).

48. Gertz, J. *et al.* Transposase mediated construction of RNA-seq libraries. *Genome Res.* **22**, 134–141 (2012).

49. Slomovic, S., Fremder, E., Staals, R. H. G., Pruijn, G. J. M. & Schuster, G. Addition of poly(A) and poly(A)-rich tails during RNA degradation in the cytoplasm of human cells. *Proc. Natl. Acad. Sci. U. S. A.* **107**, 7407–7412 (2010).

50. Ha, K. C. H., Blencowe, B. J. & Morris, Q. QAPA: a new method for the systematic analysis of alternative polyadenylation from RNA-seq data. *Genome Biol.* **19**, 45 (2018).

51. Elkon, R., Ugalde, A. P. & Agami, R. Alternative cleavage and polyadenylation: extent, regulation and function. *Nat. Rev. Genet.* **14**, 496–506 (2013).

52. Mignone, J. L., Kukekov, V., Chiang, A.-S., Steindler, D. & Enikolopov, G. Neural stem and progenitor cells in nestin-GFP transgenic mice. *J. Comp. Neurol.* **469**, 311–324 (2004).

53. Kolodziejczyk, A. A. *et al.* Single Cell RNA-Sequencing of Pluripotent States Unlocks Modular Transcriptional Variation. *Cell Stem Cell* **17**, 471–485 (2015).

54. Schubert, M., Lindgreen, S. & Orlando, L. AdapterRemoval v2: rapid adapter trimming, identification,

and read merging. *BMC Res. Notes* **9**, 88 (2016).

55. Kim, D., Langmead, B. & Salzberg, S. L. HISAT: a fast spliced aligner with low memory requirements. *Nat. Methods* **12**, 357–360 (2015).

56. O'Leary, N. A. *et al.* Reference sequence (RefSeq) database at NCBI: current status, taxonomic expansion, and functional annotation. *Nucleic Acids Res.* **44**, D733–45 (2016).

57. Harrow, J. *et al.* GENCODE: the reference human genome annotation for The ENCODE Project. *Genome Res.* **22**, 1760–1774 (2012).

58. Harrow, J. *et al.* GENCODE: producing a reference annotation for ENCODE. *Genome Biol.* **7 Suppl 1**, S4.1–9 (2006).

59. Liao, Y., Smyth, G. K. & Shi, W. The Subread aligner: fast, accurate and scalable read mapping by seed-and-vote. *Nucleic Acids Res.* **41**, e108 (2013).

60. Kroll, K. W. *et al.* Quality Control for RNA-Seq (QuaCRS): An Integrated Quality Control Pipeline. *Cancer Inform.* **13**, 7–14 (2014).

61. DeLuca, D. S. *et al.* RNA-SeQC: RNA-seq metrics for quality control and process optimization. *Bioinformatics* **28**, 1530–1532 (2012).

62. Wang, L., Wang, S. & Li, W. RSeQC: quality control of RNA-seq experiments. *Bioinformatics* **28**, 2184–2185 (2012).

63. Quinlan, A. R. & Hall, I. M. BEDTools: a flexible suite of utilities for comparing genomic features. *Bioinformatics* **26**, 841–842 (2010).

64. Kuhn, R. M., Haussler, D. & Kent, W. J. The UCSC genome browser and associated tools. *Brief. Bioinform.* **14**, 144–161 (2013).

65. Pedregosa, F. *et al.* Scikit-learn: Machine Learning in Python. *J. Mach. Learn. Res.* **12**, 2825–2830 (2011).

66. Ohio Supercomputer Center. *Columbus OH: Ohio Supercomputer Center.* http://osc.edu/ark:/19495/f5s1ph73 (1987).

**TABLES AND FIGURES**

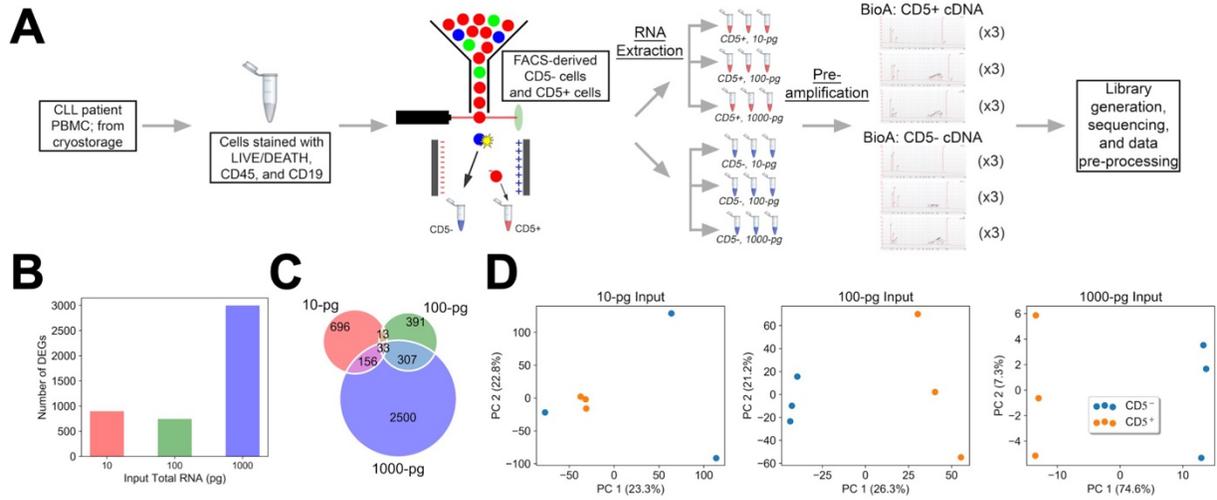

**Figure 1. lcRNA-seq analyses between sample groups without application of CLEAR. A)** Workflow for total RNA extraction and QC analysis from FACS-derived CD5+ and CD5- CLL cells as input for lcRNA-seq library generation for the development of CLEAR; **B)** The DEG counts as determined by DESeq2. Contrary to expectation, the 10-pg input has more DEGs than the 100-pg input replicates; **C)** Shared DEGs between the three input groups showing more DEGs that are unique than shared; **D)** Unsupervised analysis of CD5+ and CD5- samples by total RNA input amount. PCA reveals sample replicates separated by biological groupings at the 1,000- and the 100-pg input level but not at the 10-pg level. CLL: Chronic Lymphocytic Leukemia; PBMC: Peripheral Blood Mononuclear Cells; FACS: Fluorescence Activated Cell Sorting; BioA: Agilent Bioanalyzer RNA assay; DEGs: Differentially Expressed Genes; PC: Principal Component.

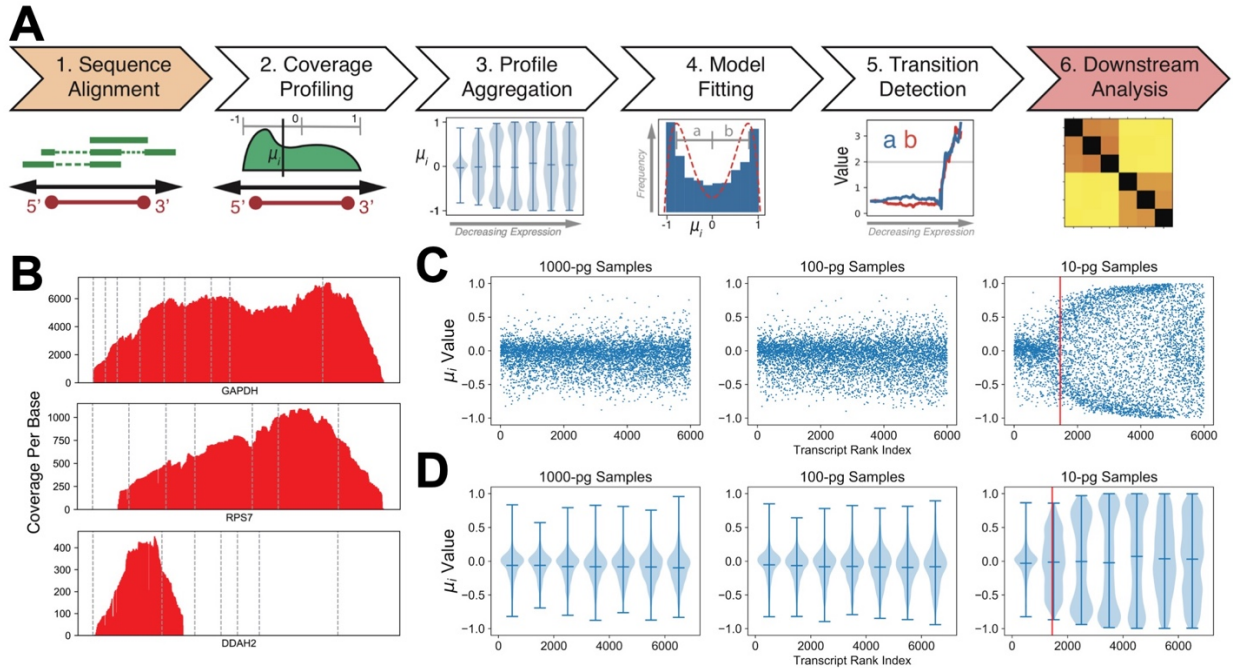

**Figure 2. CLEAR Workflow: bin-based coverage analysis by transcript expression. A)** Data analysis workflow using CLEAR to preprocess lcRNA-seq data. Step 1: Trimmed lcRNA-seq reads are aligned to the reference genome; Step 2: μ$_i$, the mean of the positional distribution of aligned reads along each individual transcript, is determined; Step 3: Transcript positional means, μ$_i$, (y-axis) are ranked and then binned by the transcript read coverage (x-axis). When μ$_i$ of a bin is ≈ 0, the read distribution is symmetrical along the length of the transcript. When μ$_i$ within a bin develops a bimodal distribution with a mode toward +1 (TTS) and -1 (TSS), its values will deviate from 0; Step 4: All available transcripts, binned into groups of 250 are fitted to a bimodal distribution model. The emergence of a bimodal distribution identifies when aggregate μ$_i$ start to deviate from a unimodal distribution around the center of the transcripts, indicated by a change in the fitting parameters a and b; Step 5: When either of the model parameters exceed a value of 2 (indicated by a gray line), transcripts beyond that point are excluded by CLEAR for differential gene expression and other downstream analysis; Step 6: CLEAR transcripts are used in downstream between-group analyses such as hierarchical clustering; **B)** example lcRNA-seq read coverage plots. Read coverage plot for GAPDH depicts a transcript with μ$_i$ ~ 0, RPS7 depicts a transcript close to the CLEAR cutoff, while DDAH2 depicts a transcript deemed too noisy by CLEAR; **C)** CLEAR profiles for 10-, 100- and 1,000-pg input mass lcRNA-seq data. The value of μ$_i$ is plotted for the 7,000 highest expressed primary transcripts for three representative samples. The red line depicts the CLEAR filtering threshold; **D)** violin plots of the same data as shown in Figure 2C. The end marks indicate the window extrema and the middle bar indicates the mean.

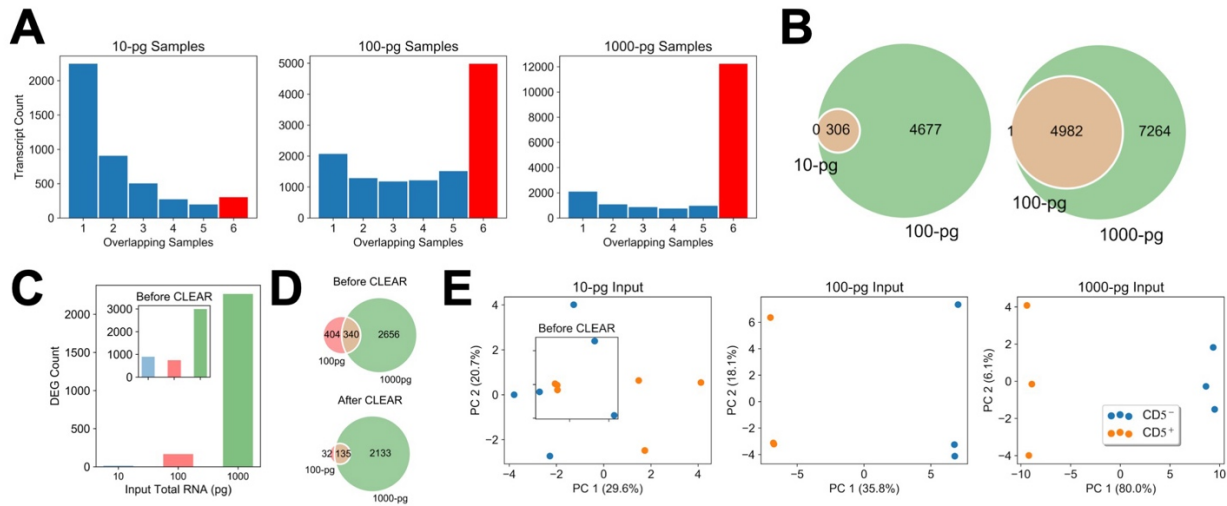

**Figure 3. lcRNA-seq analyses between sample groups after application of CLEAR. A)** CLEAR transcripts shared between samples of each input mass group. The red bars depict the number of CLEAR transcripts found in all 6 samples (replicates in both CD5+ and CD5- groups); **B)** *Left*: CLEAR transcripts overlap between 10-pg and 100-pg input mass samples; *Right*: CLEAR transcripts overlap between 100-pg and 1,000-pg input mass samples; **C)** DEG counts between CD5+ vs. CD5- cell types using only the shared CLEAR transcripts. The inset shows the data from Figure 1B without the application of CLEAR; **D)** *Bottom:* Overlap of DEGs from the 100-pg and 1,000-pg inputs (*top* repeats data from Figure 1C without CLEAR); **E)** PCA plots separating CD5+ and CD5- groups for all input masses using only CLEAR transcripts (inset repeats data from Figure 1D without CLEAR). PC: Principal Component.

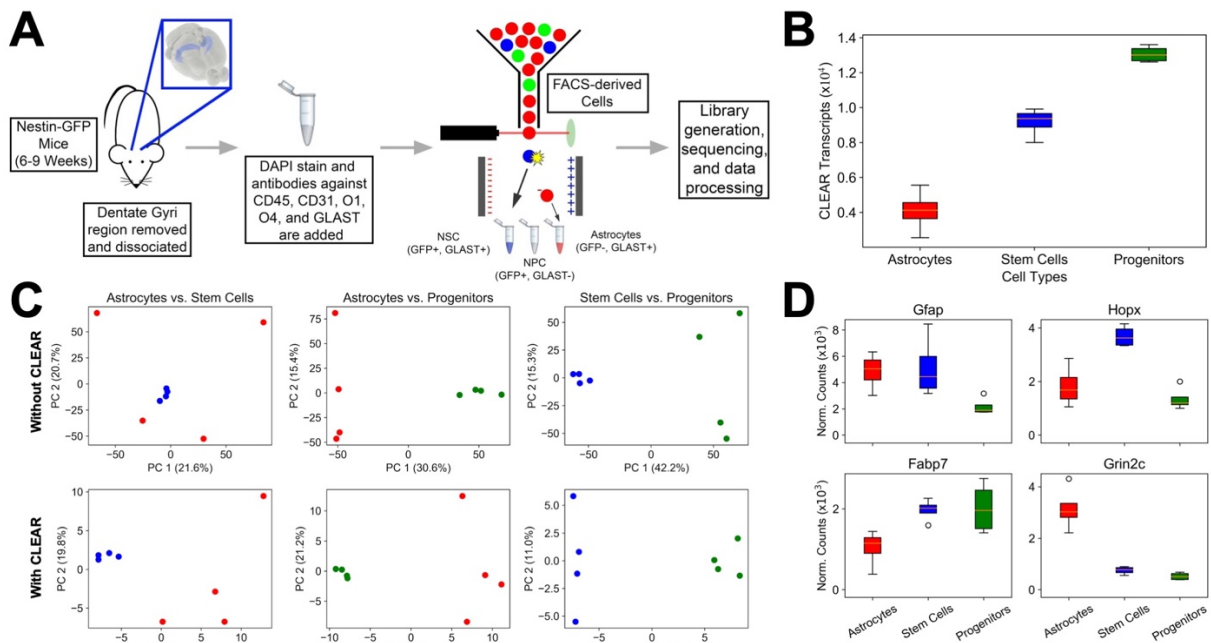

**Figure 4. Application of CLEAR to a mouse neural lcRNA-seq experiment. A)** Schematic of cell isolation and preparation for sequencing. The dentate gyri of Nestin-GFP mice were microdissected and dissociated into a single cell suspension. Cells were labeled with fluorescently conjugated antibodies

against markers for specific populations of cells present in the hippocampus. GFP+GLAST+ stem cells, GFP+GLAST- progenitor cells and GFP-GLAST+ astrocytes were isolated from live cells that were negative for microglial, oligodendroglial and endothelial markers. SMART-seq libraries were generated from these sorted cells; **B)** The means and ranges of CLEAR transcripts from each cell type (4 biological replicates per group). All groups are significantly different when compared using a t-test (p<0.01); **C)** PCA analyses by murine neuronal cell types. *Top Panels:* PCA plots using all available transcripts; *Bottom Panels:* PCA plots using only CLEAR transcripts; **D)** Normalized DESeq2 transcript counts for 4 genes that pass CLEAR and are known to be differentially expressed in murine neural stem cells, progenitors and astrocytes are used to confirm the identity of the cell populations derived from the staining and FACS strategies used to enrich these three cell populations. Boxplots: orange line, mean CLEAR transcripts for four biological replicates per neural cell type; whiskers: displaying 1.5X the inter-quartile range (IQR) beyond the first and the third quartiles; circles: outliers. FACS: Fluorescence Activated Cell Sorting; PC: Principal Component.

| Gene | Astrocyte vs. Progenitor | Progenitor vs. Stem Cell | Astrocyte vs. Stem Cell |
|---|---|---|---|
| Gfap | N.S. | ** | N.S. |
| Hopx | N.S. | **** | ** |
| Fabp7 | ** | N.S. | * |
| Grin2c | **** | N.S. | *** |
| Sox9 | F.C. | **** | F.C. |
| Neurod1 | F.C. | F.C. | F.C. |
| Dcx | F.C. | F.C. | F.C. |
| Id4 | F.C. | *** | F.C. |
| Pcna | F.C. | N.S. | F.C. |
| Mcm2 | F.C. | F.C. | F.C. |
| Ascl1 | F.C. | F.C. | F.C. |
| Eomes | F.C. | F.C. | F.C. |
| Nes | F.C. | F.C. | F.C. |
| Neurog2 | F.C. | F.C. | F.C. |

**Table 1. Application of CLEAR and DESeq2 to murine DG cell type comparisons.** Genes known to be differentially expressed in murine astrocytes, stem cells and progenitors. lcRNA-seq data were preprocessed using CLEAR prior to between-group comparisons using DESeq2. Differential gene expression analysis evaluates the effectiveness of staining and FACS strategies in enriching these three cell types. The bisecting line indicates the border between transcripts which pass CLEAR in all samples and those that do not pass CLEAR in all samples. Each comparison was processed using DESeq2 and the significance of the comparison is given. *q: FDR-corrected p-values; N.S. (Not Significant), F.C. (Failed CLEAR), * (q ≤ 0.05), ** (q ≤ 0.01), *** (q ≤ 0.001), **** (q ≤ 0.0001).*

# CLEAR: Coverage-based Limiting-cell Experiment Analysis for RNA-seq

*Logan A Walker, Michael G Sovic, Chi-Ling Chiang, Eileen Hu, Jiyeon K Denninger, Xi Chen, Elizabeth D Kirby, John C Byrd, Natarajan Muthusamy, Ralf Bundschuh, & Pearlly Yan*

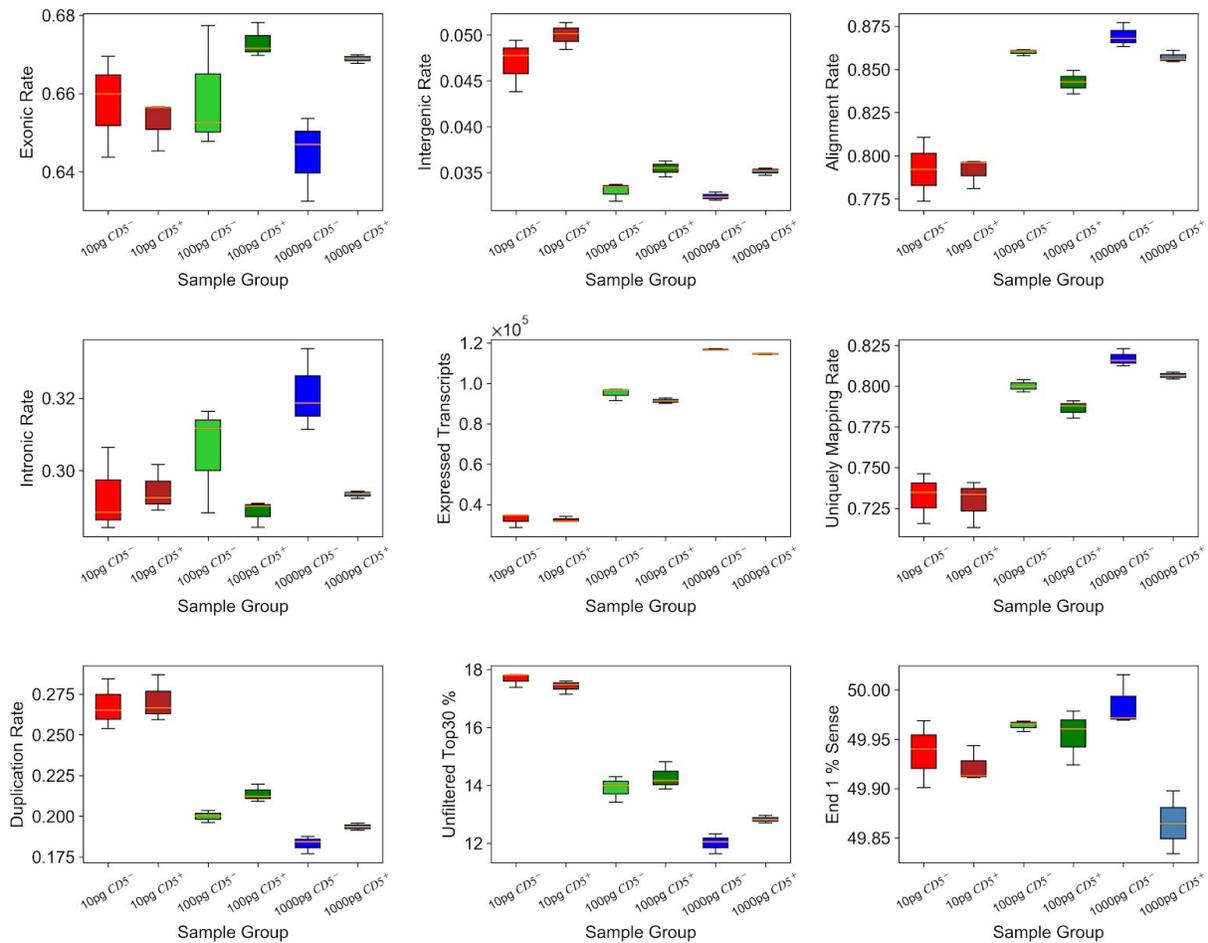

**Figure S1. Survey of quality control (QC) metrics of RNA-Seq data.** All sequencing was subject to quality control as described in **Methods**. Key metrics are summarized here. Notably, in many metrics such as Intergenic Rate, Alignment Rate, and Duplication Rate, the 10-pg groups indicate lower quality libraries than 100-pg and 1000-pg. "Top30" corresponds to the proportion of reads that belong to the 30 highest genes by expression. Boxplots: orange line, mean metric value; whiskers: displaying 1.5X the inter-quartile range (IQR) beyond the first and the third quartiles; circles: outliers.

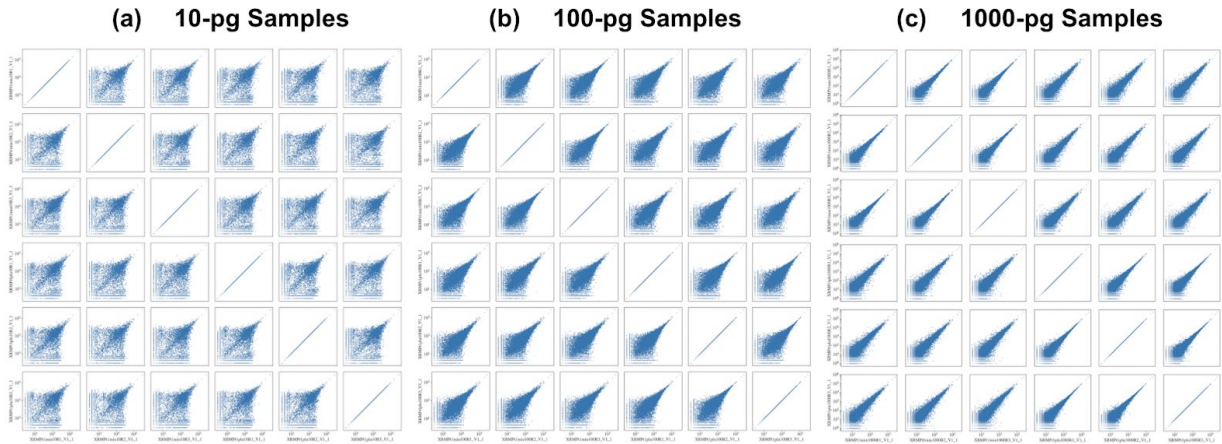

**Figure S2. Between-sample correlations of detected RNA-Seq read counts.** Scatter plots are drawn comparing each sample to each other sample for each input mass. 10-pg samples show much more scattered counts, whereas 100-pg and 1000-pg samples show progressively higher correlation.

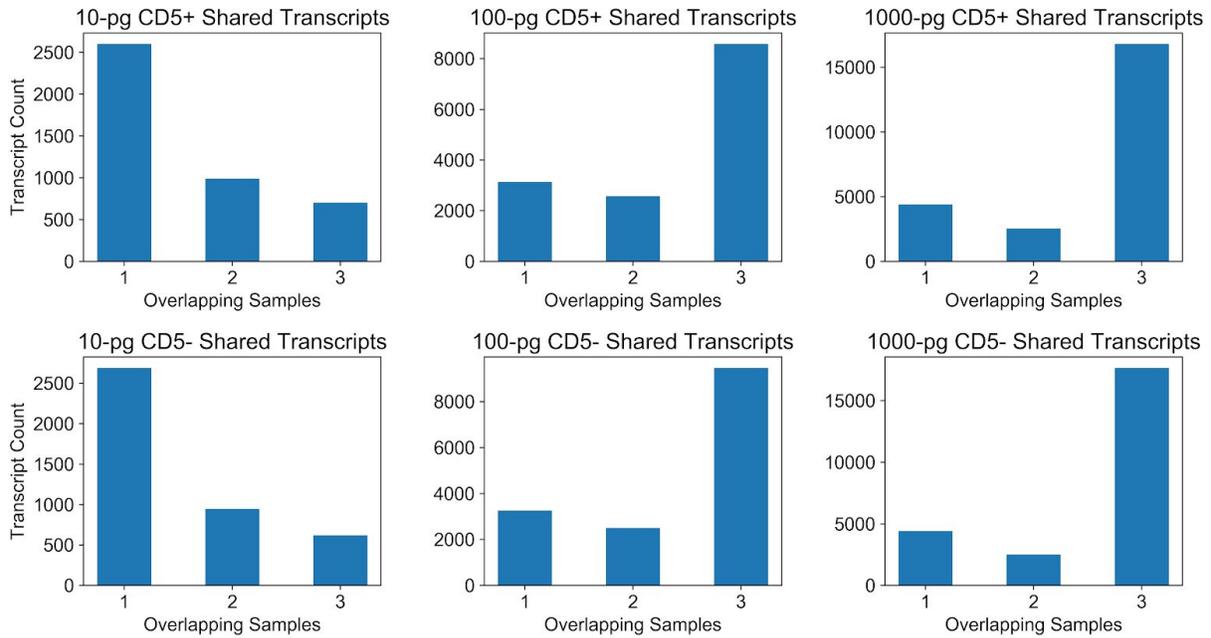

**Figure S3. Comparison of overlapping transcripts.** The analysis from Figure 3A was repeated, although CD5- and CD5+ samples were considered separately. Notably, the trend between CD5+ and CD5- mirrors that of the pooled data in Figure 3A.

Supplemental Material

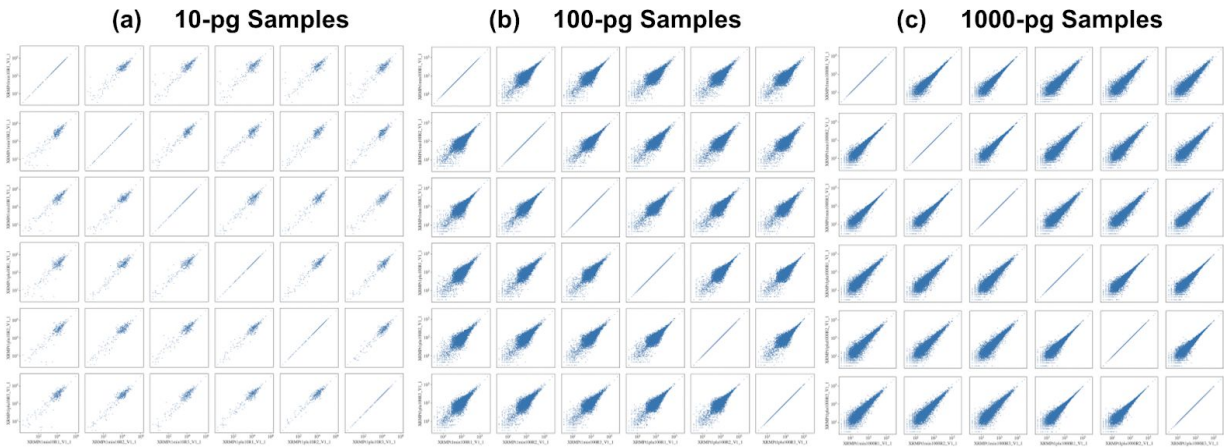

**Figure S4. CLEAR Filtering results in fewer noisy transcripts at the 10-pg sample level.** Analysis from Figure S2 was repeated using CLEAR-filtered gene counts. Notably, 10-pg samples are observed to be sparser, while the remaining data points are of much higher correlation.

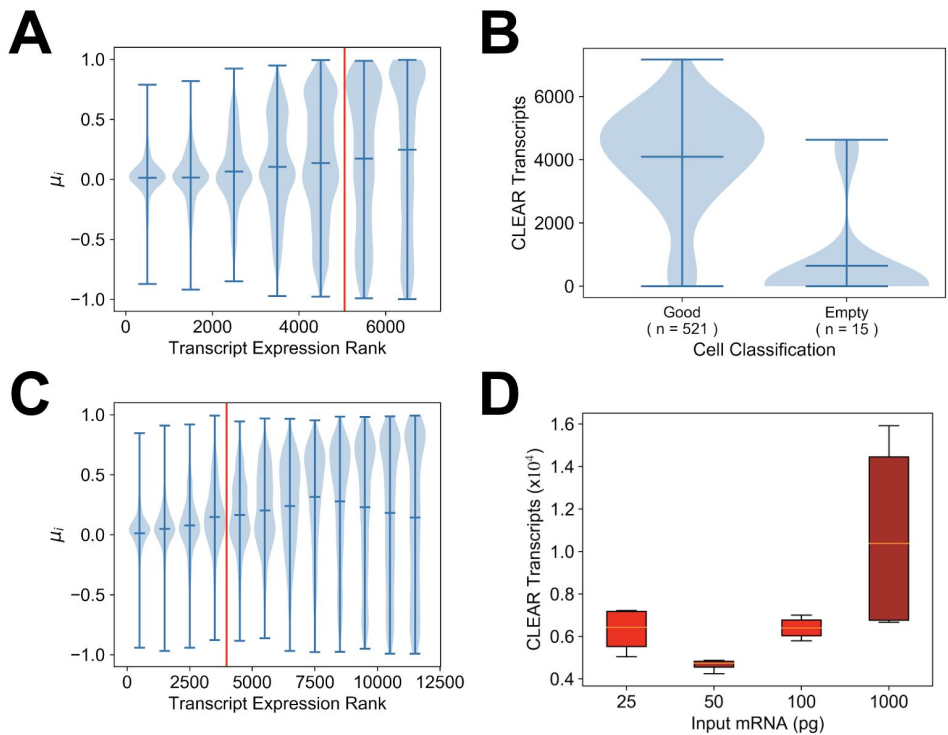

**Figure S5. Application of CLEAR to public datasets. A-B)** Data from Ilicic et al.[22] was processed using the CLEAR pipeline; **C-D)** Data from Bhargava et al.[14] was processed using the CLEAR pipeline; **A)** An example CLEAR trace from released data shows a representative separation; **B)** CLEAR transcript identity allows the separation of cells the authors classified as "Empty" from those classified as "Good." **C)** An additional example trace; **D)** CLEAR transcript counts are indicative of the input mRNA mass used to generate a sequencing library.

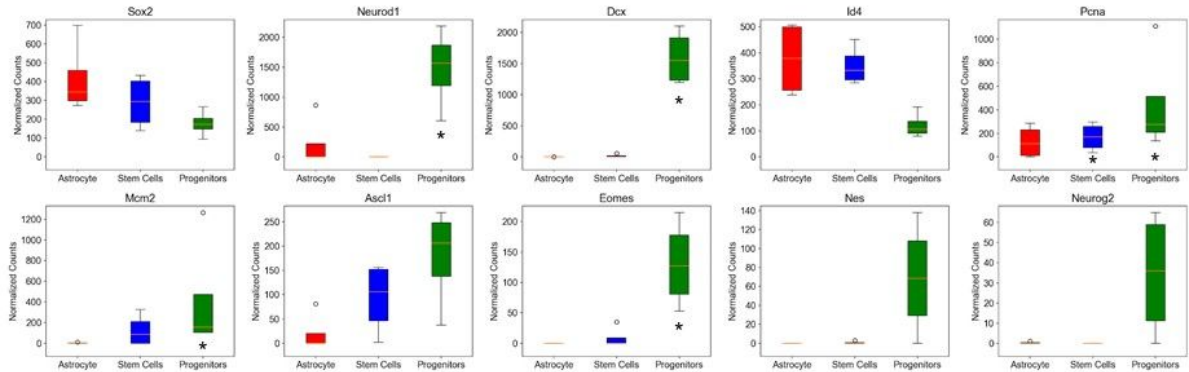

**Figure S6. Neuronal cell type markers which did not pass the CLEAR criterion.** Similar to Figure 4D, for each remaining gene, expression was plotted using the raw counts. Individual cell types which passed CLEAR filtering are indicated with an asterisk (*) below the respective box plot. Boxplots: orange line, mean CLEAR transcripts for four biological replicates per neural cell type; whiskers: displaying 1.5X the inter-quartile range (IQR) beyond the first and the third quartiles; circles: outliers.

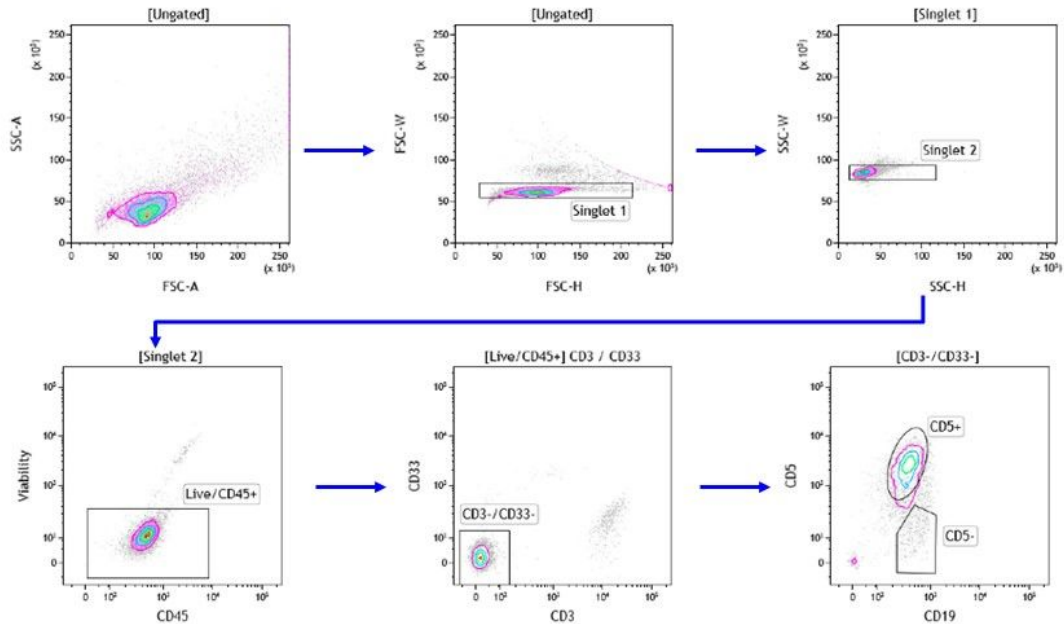

**Figure S7. FACS parameter diagrams for the enrichment of CD5+ and CD5- cells from a CLL patient PBMC sample.** FACS flow diagrams depicting the steps involved in the enrichment of CD5+ cells and CD5- cells from live CD45+, CD3-, CD33-, CD19+ cells, followed by separation into either CD5+ or CD5- collection tubes (Figure 1). FACS: Fluorescence Activated Cell Sorting.